\documentclass[11pt,a4paper]{article}
\usepackage{amsmath,amsthm,amssymb,epsfig,latexsym}
\newcommand{\be}[1]{\begin{equation}\label{#1}}
\newcommand{\ba}[1]{\begin{eqnarray}\label{#1}}
\newcommand{\ee}{\end{equation}}
\newcommand{\ea}{\end{eqnarray}}
\newcommand{\non}{\nonumber\\\rule{0pt}{30pt}}
\newcommand{\nons}{\\\rule{0pt}{30pt}}

\newcommand{\dis}{\displaystyle}
\newcommand{\eq}[1]{(\ref{#1})}

\newcommand{\Res}{\mathop{\rm Res}}

\newcommand{\Qk}{\langle\langle Q_1^\kappa\rangle\rangle}

\newcommand{\UPP}[2]{#1\hspace{-2mm}{\vphantom{\displaystyle\sum}}^{#2}}
\newtheorem{thm}{Theorem}[section]

\newtheorem{lemma}{Lemma}[section]

\def\qed{\hfill\nobreak\hbox{$\square$}\par\medbreak}
 \makeatletter
 \@addtoreset{equation}{section}
 \makeatother
 
\begin{document}
\begin{flushright}
LPENSL-TH-11/06\\
\end{flushright}
\par \vskip .1in \noindent

\vspace{24pt}

\begin{center}
\begin{LARGE}
{\bf  On correlation functions of integrable models associated to
the six-vertex R-matrix}
\end{LARGE}

\vspace{50pt}

\begin{large}

{\bf N.~Kitanine}\footnote[1]{LPTM, UMR 8089 du CNRS, Universit\'e de Cergy-Pontoise, France,
kitanine@ptm.u-cergy.fr},~~
{\bf K.~Kozlowski}\footnote[2]{ Laboratoire de Physique,  ENS Lyon,  France,
 karol.kozlowski@ens-lyon.fr},~~
{\bf J.~M.~Maillet}\footnote[3]{ Laboratoire de Physique, UMR 5672 du CNRS, ENS Lyon,  France,
 maillet@ens-lyon.fr},\\
{\bf N.~A.~Slavnov}\footnote[4]{ Steklov Mathematical Institute, Moscow, Russia, nslavnov@mi.ras.ru},~~
{\bf V.~Terras}\footnote[5]{ Laboratoire de Physique, UMR 5672 du CNRS, ENS Lyon,  France, veronique.terras@ens-lyon.fr,
on leave of absence from LPTA, UMR 5207 du CNRS, Universit\'e Montpellier II} \par

\end{large}

\vspace{80pt}

\centerline{\bf Abstract} \vspace{1cm}
\parbox{12cm}{\small%
We derive an analog of the master equation, obtained recently for correlation functions of the $XXZ$
chain, for a wide class of quantum integrable systems described by the $R$-matrix of the six-vertex model,
including in particular continuum models. This generalized master equation allows us to obtain multiple
integral representations for the correlation functions of these models. We apply this method to derive the
density-density correlation functions of the quantum non-linear Schr\"odinger model.}
\end{center}

\newpage

\section{Introduction}

The method to study the spectrum of the Heisenberg
antiferromagnet was proposed by H. Bethe in 1931 \cite{Bet31}.
This method, called later Bethe ansatz, was found to be very
effective for a complete description of the spectrum of quantum
solvable models. During the ensuing years the Bethe ansatz was
successfully extended and applied to the study of various problems
(see e.g. \cite{Orb58,Wal59,GauL83,LieSM61,LieM66,YanY66}).

A further development of this method was done in \cite{FadST79},
where the algebraic version of the Bethe ansatz was formulated. In
the algebraic Bethe ansatz one deals with a set of operators (the
operators entries of the quantum monodromy matrix) possessing
quadratic commutation relations (governed by the $R$-matrix
solving the Yang-Baxter equation). Different representations of
this algebra correspond to different physical models associated to
the same $R$-matrix. Thus, the algebraic Bethe ansatz allows one to
consider a wide class of quantum integrable models from a uniform
algebraic viewpoint (see \cite{BogIK93L} and references therein).

In spite of its successful application to the
calculation of the spectrum of quantum Hamiltonians, the
possibility to use the Bethe ansatz to evaluate the corresponding
correlation functions remained involved for a long time. Basically
the first complete results for integrable models correlation
functions concerned the models equivalent to free fermions
\cite{LieSM61,Mcc68,Len66,SatMJ78,KorS89,ColIKT93} for which
different methods apply. Attempts to compute the correlation
functions of local operators by the algebraic Bethe ansatz for
more general models were undertaken in
\cite{IzeK84,IzeK85,Kor87,KojKS97}. One of the main problems there
is to embed in an appropriate way the local operators into the
quadratic Yang-Baxter algebra describing the spectrum of these
models.

The solution of the quantum inverse scattering problem, first obtained for the $XXZ$ chain \cite{KitMT99},
leads to a neat solution to this question: the local operators are then given as explicit products of the
quantum monodromy matrix entries. Hence it  gives a powerful tool for the computation of the correlation
functions via the algebraic Bethe ansatz (see e.g. \cite{KitMT00} and further developments reviewed in
\cite{KitMST05c}). In turn, the explicit and compact expressions for local operators in terms of the
entries of the quantum monodromy matrix reduce finally the calculation of the correlation functions to the
well studied problem of the computation of scalar products in the algebraic Bethe ansatz framework
\cite{Sla89,Sla95,KitMT99}. It is worth mentioning, however, that the solution of the quantum inverse
scattering problem has been obtained so far only for lattice models \cite{MaiT00}, the case of continuum
(field theory) models remaining an open question.

The main goal of this paper is  to extend some results obtained
for the $XXZ$ chain to a wider class of integrable models
described by the algebraic Bethe ansatz, but for which the
explicit solution of the quantum inverse scattering problem is not
known at present.  Namely, we derive an analog of the master
equation for correlation functions of the $XXZ$ chain obtained in
\cite{KitMST05a} for the generalized model proposed in
\cite{Kor82,IzeK84,IzeK85}. This model includes as a particular case
continuum (field theory) models, the quantum Non Linear
Schr\"odinger (NLS) model being one typical example. Just like in
the case of $XXZ$ chain, this master equation can be used to
derive multiple integral representations for a certain class of
correlation functions. In particular, we obtain the
density-density correlation functions of the quantum NLS model.
The basic idea to reach this goal is to find a way to embed
certain local operators of these models into the Yang-Baxter
algebra of the quantum monodromy matrix entries, or at least, into
the algebra of the quantum monodromy matrix entries corresponding
to a subpart of the model. This can be achieved in particular for
the number of particle operator associated to a segment $[0,x]$ in
the quantum NLS model. Using then algebraic Bethe ansatz
techniques, together with the determinant formula for the scalar
products of Bethe states \cite{Sla89,Sla95}, it is possible to
compute the corresponding correlation functions in a rather
compact way, hence leading to density-density correlation
functions. The same technique can be applied for more general
correlation functions as well.

The content of the paper is as follows. In Section 2 we recall some basic definitions and formulas
necessary to study the generalized model in the framework of algebraic Bethe ansatz. In Section 3 we
consider the two-site generalized model and derive the master equation for the expectation value of the
operator of number of particles in the first site. In Section 4 we use this master equation to evaluate
the multiple integral representation for the density-density correlation function of the Quantum Nonlinear
Schr\"odinger model. The proofs of some algebraic identities are collected in three appendices.


\section{Generalized model}\label{BA}

In this section we recall briefly the main definitions of the algebraic Bethe ansatz and introduce
necessary notations. We refer the reader to \cite{FadST79}, \cite{BogIK93L} for more details.

The main objects of the algebraic Bethe ansatz are the $R$-matrix $R(\lambda)$, the monodromy matrix
$T(\lambda)$ and the pseudovacuum vector $|0\rangle$. We consider quantum
integrable systems associated to the $R$-matrix of the six-vertex model\footnote[1]{%
One can also consider the particular case where hyperbolic functions in the $R$-matrix  degenerate into
the rational ones.}
\be{R}
 R(\lambda)=\left(
\begin{array}{cccc}
\sinh(\lambda+\eta)&0&0&0\\
0&\sinh\lambda&\sinh\eta&0\\
0&\sinh\eta&\sinh\lambda&0\\
0&0&0&\sinh(\lambda+\eta)
\end{array}\right),
\ee
where $\lambda$ and $\eta$ are complex numbers. Usually the parameter $\eta$ is related to the coupling
constant of the Hamiltonian.

The monodromy matrix of such model  is  a $2\times 2$ matrix
\be{ABAT} T(\lambda)=\left(
\begin{array}{cc}
A(\lambda)&B(\lambda)\\
C(\lambda)&D(\lambda)
\end{array}\right)
\ee
with operator-valued entries $A,B,C$ and $D$ which depend on a complex parameter $\lambda$ and act in the
quantum space of states of the Hamiltonian. These operators satisfy a set of quadratic commutation
relations given by
\be{RTT}
 R(\lambda-\mu)\UPP{T}{1}(\lambda)\UPP{T}{2}(\mu)=
  \UPP{T}{2}(\mu)\UPP{T}{1}(\lambda)R(\lambda-\mu).
  \ee
 Here $\UPP{T}{1}(\lambda)=T(\lambda)\otimes I$ and $\UPP{T}{2}(\lambda)=
 I\otimes T(\lambda)$.

 The pseudovacuum $|0\rangle$ is characterized by the action
 of the entries of the monodromy matrix on this vector
 \be{action}
 A(\lambda)|0\rangle=a(\lambda)|0\rangle,\qquad
  D(\lambda)|0\rangle=d(\lambda)|0\rangle,\qquad
   C(\lambda)|0\rangle=0.
\ee
Here $a(\lambda)$ and $d(\lambda)$ are complex functions depending
on the specific model. In the framework of the generalized model
they remain free functional parameters. Various models associated
to the same $R$-matrix  are described by different representations
of the  ratio $r(\lambda)=a(\lambda)/d(\lambda)$.

The dual pseudovacuum $\langle0|=|0\rangle^\dagger$,
 $\langle0|0\rangle=1$ is characterized by similar equations
 \be{daction}
 \langle0|A(\lambda)=a(\lambda)\langle0|,\qquad
  \langle0|D(\lambda)=d(\lambda)\langle0|,\qquad
   \langle0|B(\lambda)=0.
\ee

The trace of the monodromy matrix ${\cal T}(\mu)=A(\mu)+D(\mu)$
appears to be the generating function of the integrals of motion
of the model. In particular the eigenstates of ${\cal T}(\mu)$
coincide with the eigenstates of the Hamiltonian. We shall also
consider a more general object
\be{Twis-T-M} {\cal T}_\kappa(\mu)=A(\mu)+\kappa D(\mu), \ee
which is called the twisted transfer matrix. Here $\kappa$ is a
complex parameter. The eigenstates (and their dual ones) of the
operator ${\cal T}_\kappa(\mu)$ can be written in the form
\be{ABAES}
 |\psi_\kappa(\{\lambda\})\rangle=
 \prod_{j=1}^{N}B(\lambda_j)|0\rangle,
 \qquad
 \langle\psi_\kappa(\{\lambda\})|=
 \langle 0|\prod_{j=1}^{N}C(\lambda_j),
 \qquad N=0,1,\dots
\ee
where  the parameters $\lambda_1,\dots,\lambda_N$ satisfy the system of twisted Bethe equations
 \be{TTMBE_Y}
 {\cal Y}_\kappa(\lambda_j|\{\lambda\})=0, \qquad
 j=1,\dots,N.
 \ee
Here
\be{TTM_Y-def} {\cal Y}_\kappa(\mu|\{\lambda\}) = a(\mu)\prod_{k=1}^{N}\sinh(\lambda_k-\mu+\eta) +
\kappa\,d(\mu)
 \prod_{k=1}^{N}\sinh(\lambda_k-\mu-\eta).
 \ee
The corresponding eigenvalue of ${\cal T}_\kappa(\mu)$ on
$|\psi_\kappa(\{\lambda\})\rangle$ (or on a dual eigenstate)
reads
\be{ABAEV} \tau_\kappa(\mu|\{\lambda\})=
a(\mu)\prod_{k=1}^{N}\frac{\sinh(\lambda_k-\mu+\eta)}{\sinh(\lambda_k-\mu)} + \kappa\,d(\mu)
\prod_{k=1}^{N}\frac{\sinh(\mu-\lambda_k+\eta)}{\sinh(\mu-\lambda_k)}. \ee
Formulas \eq{ABAES}--\eq{ABAEV} remain valid for the particular
case $\kappa=1$ for which we omit  the subscript $\kappa$, denoting for
example, $|\psi(\{\lambda\})\rangle$, ${\cal Y}(\mu|\{\lambda\})$
and $\tau(\mu|\{\lambda\})$.

A central role in the calculation of the correlation functions is
played by scalar products of states
\be{defSN}
 S_N(\{\mu\}|\{\lambda\})=
 \langle 0|\prod_{j=1}^{N}C(\mu_j)
 \prod_{j=1}^{N}B(\lambda_j)|0\rangle,
\ee
where the parameters $\mu_1,\dots,\mu_N$ and
$\lambda_1,\dots,\lambda_N$ are arbitrary complex numbers and
generically do not satisfy any constrains. Using the commutation
relations \eq{RTT} one can obtain an explicit, but not very
compact formula for this quantity \cite{IzeK84,Ize87,BogIK93L}.
 \begin{multline}\label{Gen-case}
 \langle0|\prod_{j=1}^N\frac{C(\mu_j)}{d(\mu_j)}
 \prod_{j=1}^N\frac{B(\lambda_j)}{d(\lambda_j)}|0\rangle=\prod_{j>k}^N
 \frac1{\sinh(\lambda_j-\lambda_k)\sinh(\mu_k-\mu_j)}\nons
 \times\sum_{\alpha\cup\bar\alpha\atop{\gamma\cup\bar\gamma}}(-1)^{[P(\alpha)]+[P(\gamma)]}
 \prod_{a\in\gamma}r(\mu_a)\prod_{a\in\bar\alpha}r(\lambda_a)
 \det_{j\in\alpha\atop{k\in\gamma}}t(\lambda_j,\mu_k)
 \det_{j\in\bar\alpha\atop{k\in\bar\gamma}}t(\mu_k,\lambda_j)\nons
 \times
 \prod_{a\in\gamma}\prod_{b\in\alpha}h(\lambda_b,\mu_a)
 \prod_{a\in\bar\alpha}\prod_{b\in\bar\gamma}h(\mu_b,\lambda_a)
 \prod_{a\in\bar\alpha}\prod_{b\in\alpha}h(\lambda_b,\lambda_a)
 \prod_{a\in\gamma}\prod_{b\in\bar\gamma}h(\mu_b,\mu_a),
 \end{multline}
Let us describe the notations used in \eq{Gen-case}. First, we have introduced the two functions
 \be{ht}
 h(\lambda,\mu)=\sinh(\lambda-\mu+\eta),\qquad
 t(\lambda,\mu)=\frac{\sinh\eta}{\sinh(\lambda-\mu)\sinh(\lambda-\mu+\eta)}.
 \ee
The set of the parameters $\{\lambda\}$ is divided  into two disjoint subsets
$\{\lambda\}=\{\lambda\}_{\alpha}\cup\{\lambda\}_{\bar\alpha}$. Similarly
$\{\mu\}=\{\mu\}_{\gamma}\cup\{\mu\}_{\bar\gamma}$. Hereby $\#\{\lambda\}_{\alpha}= \# \{\mu\}_{\gamma}$.
The parameters in each subset are ordered in the natural way, namely
$\{\lambda_{\alpha_1},\lambda_{\alpha_2},\dots\}$ if $\alpha_1<\alpha_2<\dots$ etc. The sum in
\eq{Gen-case} is taken with respect to all possible partitions of this kind. The symbol $P(\alpha)$
(respectively $P(\gamma)$) means the permutation $P(\{\alpha\},\{\bar\alpha\})=1,\dots,N$ (respectively
$P(\{\gamma\},\{\bar\gamma\})=1,\dots,N$). Similar notations will be used through all the paper.

The formula \eq{Gen-case} can be drastically simplified if one of
the states is an eigenstate of the twisted transfer-matrix. Let us
introduce, for arbitrary positive integers $n$ and $n'$ ($n\le n'$)
and
arbitrary sets of variables $\lambda_1,\dots,\lambda_n$, $\mu_1,%
\dots,\mu_n$ and $\nu_1,\dots,\nu_{n'}$, such that $\{\lambda\} \subset \{\nu\}$,  the following $n\times
n$ matrix $\Omega_\kappa(\{\lambda\},\{\mu\}|\{\nu\})$
\begin{multline} \label{matH}
  (\Omega_\kappa)_{jk}(\{\lambda\},\{\mu\}|\{\nu\})=
  a(\mu_k)\,t(\lambda_j,\mu_k)\,\prod_{a=1}^{n'} \sinh(\nu_a-\mu_k+\eta)\\
   -\kappa\, d(\mu_k)\,t(\mu_k,\lambda_j)\,\prod_{a=1}^{n'} \sinh(\nu_a-\mu_k-\eta).
\end{multline}
Then, if $\{\lambda\}$ satisfies the system of twisted Bethe equations,
 \begin{align}\label{Sc-prod}
 \langle0|\prod_{j=1}^N C(\mu_j)|\psi_\kappa(\{\lambda\})\rangle &=
 \langle\psi_\kappa(\{\lambda\})|\prod_{j=1}^N B(\mu_j)|0\rangle\\
 &=\frac{\prod_{j=1}^N d(\lambda_j)}{\prod_{j>k}^N\sinh(\lambda_j-\lambda_k)\sinh(\mu_k-\mu_j)}
 \cdot \det_N \Omega_\kappa(\{\lambda\},\{\mu\}|\{\lambda\}).
 \end{align}
Different proofs of this formula for the case $\kappa=1$ were
given in \cite{Sla89,Sla95}, and for the particular case of the
$XXZ$ chain in \cite{KitMT99}. For completeness we give in
Appendix A another proof for arbitrary $\kappa$, which will be
used later.

Observe that formula \eq{Sc-prod} implies two different representations for the particular case of the
scalar product $\langle\psi(\{\mu\})|\psi_\kappa(\{\lambda\})\rangle$, for which in addition $\{\mu\}$ is
a solution of ordinary (non-twisted) Bethe equations:
 \begin{align}\label{Sc-prod-1}
 \langle\psi(\{\mu\})|\psi_\kappa(\{\lambda\})\rangle&=
 \frac{\prod_{j=1}^N d(\lambda_j)}{\prod_{j>k}^N\sinh(\lambda_j-\lambda_k)\sinh(\mu_k-\mu_j)}
 \cdot \det_N \Omega_\kappa(\{\lambda\},\{\mu\}|\{\lambda\}) \\
 &=\frac{\prod_{j=1}^N d(\mu_j)}{\prod_{j>k}^N\sinh(\lambda_j-\lambda_k)\sinh(\mu_k-\mu_j)}
 \cdot \det_N \Omega(\{\mu\},\{\lambda\}|\{\mu\}),
 \end{align}
where $\Omega=\left.\Omega_\kappa\right|_{\kappa=1}$. Indeed, one can consider  either the state
$\langle\psi(\{\mu\})|$ or the state $|\psi_\kappa(\{\lambda\})\rangle$ as a particular case of an
arbitrary state. We conclude that
 \be{OOk}
 \prod_{j=1}^N d(\lambda_j)\cdot\det_N \Omega_\kappa(\{\lambda\},\{\mu\}|\{\lambda\})=
 \prod_{j=1}^N d(\mu_j)\cdot\det_N \Omega(\{\mu\},\{\lambda\}|\{\mu\}).
 \ee
This identity holds if the parameters $\{\lambda\}$ (respectively $\{\mu\}$) are solutions of twisted
Bethe equations \eq{TTMBE_Y} (respectively ordinary Bethe equations with $\kappa=1$), see Appendix
\ref{id-det} for a direct proof. Otherwise both sides of the equation \eq{OOk} define different functions
of $\{\lambda\}$ and $\{\mu\}$.

Setting $\mu_j=\lambda_j$, $j=1,\dots,N$  in \eq{Sc-prod} one has
 \be{norm}
 \langle \psi_\kappa(\{\lambda\})|\psi_\kappa(\{\lambda\})\rangle = \frac{\prod_{a=1}^{N}
 d(\lambda_a)} {\prod\limits_{a,b=1\atop{a\ne b}}^N\sinh(\lambda_a-\lambda_b)} \cdot
 \det_N\left(-\frac\partial{\partial\lambda_k} {\cal Y}_\kappa(\lambda_j|\{\lambda\})\right),
 \ee
where ${\cal Y}_\kappa$ is given by \eq{TTM_Y-def}.


\section{Two-site model}

A number of correlation functions can be computed in the framework of the generalized two-site model
introduced in \cite{IzeK84,IzeK85} (see also \cite{BogIK93L}). In the two-site model the monodromy matrix
$T(\lambda)$ is presented as the product of two operators $T(\lambda)=T_2(\lambda)T_1(\lambda)$, where
 \be{ABAT-2s}
 T_j(\lambda)=\left(
\begin{array}{cc}
A_j(\lambda)&B_j(\lambda)\\
C_j(\lambda)&D_j(\lambda)
\end{array}\right), \qquad j=1,2.
\ee
The commutation relations between the entries of each $T_j(\lambda)$ are given by the $R$-matrix \eq{R},
the operators from the different sites commuting with each other. The monodromy matrices $T_j(\lambda)$
have their own pseudovacuum vector $|0\rangle_j$ and dual one $\langle0|_j$. Hereby
$|0\rangle=|0\rangle_2\otimes|0\rangle_1$. The action of the operators on these vectors is similar to
\eq{action}, \eq{daction}
 \be{action-2s}
 \begin{array}{l}
 {\dis
 A_j(\lambda)|0\rangle_j=a_j(\lambda)|0\rangle_j,\qquad
  D_j(\lambda)|0\rangle=d_j(\lambda)|0\rangle_j,\qquad
   C_j(\lambda)|0\rangle_j=0,}\non
{\dis \langle0|_jA_j(\lambda)=a_j(\lambda)\langle0|_j,\qquad
  \langle0|_jD_j(\lambda)=d_j(\lambda)\langle0|_j,\qquad
   \langle0|_jB_j(\lambda)=0.}\end{array}
\ee
It is easy to see that $a(\lambda)=a_1(\lambda)a_2(\lambda)$ and $d(\lambda)=d_1(\lambda)d_2(\lambda)$.
Below we shall also use the functions
 \be{ff}
 l(\lambda)=\frac{a_1(\lambda)}{d_1(\lambda)},\qquad m(\lambda)=\frac{a_2(\lambda)}{d_2(\lambda)},
 \qquad r(\lambda)=l(\lambda)m(\lambda).
 \ee

Following the work \cite{IzeK84} we introduce now the operator $Q_1$  of number of particles  in the first
site. This operator is defined by its commutation relations with the entries of the monodromy matrices
$T_j(\lambda)$
 \be{Q1-comm}
 [Q_1,B_1(\lambda)]=B_1(\lambda), \qquad [Q_1,C_1(\lambda)]=-C_1(\lambda).
 \ee
All other commutation relations are trivial. The action of $Q_1$ on the pseudovacuum is given by
 \be{actQ1}
 Q_1|0\rangle_1=0, \qquad  \langle0|_1Q_1=0.
 \ee

The expectation value of the operator $e^{\beta Q_1}$, where $\beta$ is a complex number, is a generating
function for correlation functions. In order to compute this expectation value  in the framework of the
generalized two-site model one should use the commutation relations \eq{Q1-comm}. We use the result
obtained in \cite{IzeK84,IzeK85}. Let
 \be{int-mat-el}
 \Qk=\langle0|\prod_{j=1}^N\frac{C(\mu_j)}{d(\mu_j)}\cdot e^{\beta Q_1}
 \cdot\prod_{j=1}^N\frac{B(\lambda_j)}{d(\lambda_j)}|0\rangle\cdot\prod_{j>k}^N
 \sinh(\lambda_j-\lambda_k)\sinh(\mu_k-\mu_j),\qquad \kappa=e^\beta,
\ee
where $\{\lambda\}$ and $\{\mu\}$ are generic complex numbers. Then
\begin{multline}\label{IK}
 \Qk=\sum_{\alpha\cup\bar\alpha\atop{\gamma\cup\bar\gamma}}
 \kappa^{n}
\langle0|_1\prod_{j\in\gamma}^N\frac{C_1(\mu_j)}{d_1(\mu_j)}
 \prod_{j\in\alpha}^N\frac{B_1(\lambda_j)}{d_1(\lambda_j)}|0\rangle_1
 \langle0|_2\prod_{j\in\bar\gamma}^N\frac{C_2(\mu_j)}{d_2(\mu_j)}
 \prod_{j\in\bar\alpha}^N\frac{B_2(\lambda_j)}{d_2(\lambda_j)}|0\rangle_2\nons
\times\prod_{a>b\atop{a,b\in\alpha}}\sinh(\lambda_a-\lambda_b)
\prod_{a<b\atop{a,b\in\gamma}}\sinh(\mu_a-\mu_b)
\prod_{a>b\atop{a,b\in\bar\alpha}}\sinh(\lambda_a-\lambda_b)
\prod_{a<b\atop{a,b\in\gamma}}\sinh(\mu_a-\mu_b) \nons
\times (-1)^{[P(\alpha)]+[P(\gamma)]}\prod_{a\in\alpha}m(\lambda_a)\prod_{a\in\bar\gamma}l(\mu_a)
\prod_{a\in\alpha}\prod_{b\in\bar\alpha}
 h(\lambda_b,\lambda_a)
 \prod_{a\in\gamma}\prod_{b\in\bar\gamma}
 h(\mu_a,\mu_b).
 \end{multline}
Here $n$ denotes the number of elements in the subset $\{\alpha\}$ (or $\{\gamma\}$). Thus, we see that
this matrix element can be reduced to the scalar products in the first and in the second sites. For both
of them one can use the representation \eq{Gen-case} with appropriate replacement of $r(\lambda)$ by
$l(\lambda)$ or $m(\lambda)$.

Substituting \eq{Gen-case} for each scalar product in \eq{IK} we obtain a sum over $4$ partitions of the
set $\{\lambda\}$ and $4$ partitions of the set $\{\mu\}$. It was shown in the papers \cite{IzeK84,
IzeK85} that the sum over four of these partitions can be calculated, if the sets of the parameters
$\{\lambda\}$ and $\{\mu\}$ are two different solutions of Bethe equations. This result, however, does not
allow one to set $\{\lambda\}=\{\mu\}$ in this simplified formula, since two different solutions of Bethe
equations can not be sent to each other. We consider therefore a more general case, when only the set of the
parameters $\{\lambda\}$ satisfies Bethe equations, while the parameters $\{\mu\}$ remain arbitrary
complex numbers. In a sense this case is analogous to the one considered in Section \ref{BA} for scalar
products.

The main result of this section and of the paper is the
generalized master equation for the matrix element of the operator
$e^{\beta Q_1}$.

\begin{thm}\label{M-theor}
Let the parameters $\lambda_1,\dots,\lambda_N$ satisfy Bethe equations at $\kappa=1$, while
$\mu_1,\dots,\mu_N$ are generic complex numbers. Then
\begin{multline}\label{Mult-int}
 \Qk=\frac1{N!}\oint\limits_{\Gamma(\{\mu\}\cup\{\lambda\})}\prod_{j=1}^N
 \frac{dz_j}{2\pi i}\prod_{j=1}^N\frac{l(z_j)l^{-1}(\lambda_j)d(z_j)}
 {d(\lambda_j)d(\mu_j)}\nons
 \times
 \frac{\det_N\Omega_\kappa(\{z\},\{\mu\}|\{z\})\det_N\Omega_\kappa(\{z\},\{\lambda\}|\{z\})}
 {\prod\limits_{j=1}^N{\cal Y}_\kappa(z_j|\{z\})},
 \end{multline}
where the integration contour $\Gamma(\{\mu\}\cup\{\lambda\})$
surrounds the poles of $\Omega_\kappa$ in the points $\{\lambda\}$
and $\{\mu\}$ and no other pole of the integrand.
\end{thm}

The proof is given in Appendix \ref{PT}. We would like to
stress here that the only information used to derive this formula
are the commutation relations \eq{Q1-comm}.

{\sl Note.} In the $XXZ$ model the two ``sites'' of the generalized model
correspond to the division of the chain into two subchains.
In this framework, the operator $e^{\beta Q_1}$ has an explicit representation in terms of
local spin operators,
 \be{GFdefQ}
  e^{\beta Q_{1}}=\prod_{n=1}^m\left(\frac{1+\kappa}2+\frac{1-\kappa}2\cdot \sigma_n^z\right)=
 \exp\left(\frac\beta2\sum_{n=1}^{m}(1-\sigma_n^z)\right),\qquad e^\beta=\kappa,
  \ee
where $m$ is the length of the first subchain. Using the solution of the inverse scattering problem
\cite{KitMT00} one can obtain the equation \eq{Mult-int} by the method described in \cite{KitMST05a}.
However, this approach fails in the case of the generalized model, since the solution of the inverse
scattering problem for such a model is not known at present.

Since the parameters $\{\mu\}$ in the equation \eq{Mult-int} do not satisfy any constraints one can
consider the particular case $\mu_j=\lambda_j$, $j=1,\dots,N$. In this case we obtain for the normalized
expectation value of the operator $e^{\beta Q_1}$
\begin{multline}\label{GME-1}
 \langle e^{\beta Q_1}\rangle
=\frac{\langle\psi(\{\lambda\})|e^{\beta Q_1}
|\psi(\{\lambda\})\rangle}{\langle\psi(\{\lambda\})|\psi(\{\lambda\})\rangle}
 =\frac{(-1)^N}{N!}\oint\limits_{\Gamma(\{\lambda\})}\prod_{j=1}^N
 \left(\frac{dz_j}{2\pi i}\cdot\frac{l(z_j)d(z_j)}
 {l(\lambda_j)d(\lambda_j)}\right)\nons
  \times
 \frac{\Bigl[\det_N\Omega_\kappa(\{z\},\{\lambda\}|\{z\})\Bigr]^2}
  {\prod\limits_{j=1}^N{\cal Y}_\kappa(z_j|\{z\})
 \det_N\frac{\partial{\cal Y}(\lambda_j|\{\lambda\})}{\partial\lambda_k}}.
 \end{multline}
%

\section{Quantum Nonlinear Schr\"odinger equation}

As an application of the master equation \eq{GME-1} we derive now
a multiple integral representation for the density-density correlation
function of the one-dimensional Bose gas. The method described below
is very similar to the one used in \cite{KitMST05a,KitMST05b} for
the $XXZ$ chain, therefore we refer the reader to the mentioned
papers for more details.

The model of the one-dimensional Bose-gas is described by the Quantum Nonlinear Schr\"odinger equation.
The Hamiltonian of this model on the ring of length $L$ is given by
 \be{Ham}
 H=\int_0^L\Bigl(\partial_x\psi^\dagger(x)\partial_x\psi(x)+c\psi^\dagger(x)\psi^\dagger(x)\psi(x)\psi(x)-h\psi^\dagger(x)\psi(x)\Bigr)\,dx,
 \ee
 where $c>0$ is the coupling constant and $h>0$ the chemical potential. The operators $\psi^\dagger$ and
 $\psi$ are canonical Bose-fields obeying standard equal-time commutation relations. This model was
 studied via the algebraic Bethe ansatz in \cite{Kor82,IzeK84,BogIK93L}. The operator $Q_1$ giving the number of particle in the interval $[0,x]$ for this model has the
 following explicit representation
  \be{Q1NS}
  Q_1=\int_0^x \psi^\dagger(z)\psi(z)\,dz.
  \ee
The commutation relations \eq{Q1-comm} can be proven directly.
Then, the ground state expectation value of $e^{\beta Q_1}$ can be
used to compute the density-density correlation
function \cite{IzeK84}
 \be{dd}
 \langle \psi^\dagger(x)\psi(x)\cdot \psi^\dagger(0)\psi(0)\rangle=
 \frac12\frac{\partial^2}{\partial\beta^2}\left.\frac{\partial^2}{\partial x^2}
 \langle e^{\beta Q_1}\rangle\right|_{\beta=0}.
 \ee
The master equation \eq{GME-1} permits us to obtain in a simple way a multiple integral representation for
this quantity in the thermodynamic limit.

In the case of the Quantum Nonlinear Schr\"odinger  equation one has
\be{rlm}
 a(\lambda)=e^{-iL\lambda/2},\qquad d(\lambda)=e^{iL\lambda/2},
 \qquad l(\lambda)=e^{-ix\lambda},
 \ee
One should also replace all hyperbolic functions in $\Omega_\kappa$ and ${\cal Y}_\kappa$ by the rational
ones and set $\eta=-ic$. The system of Bethe equations then reads
\be{NS-BE}
 e^{iL\lambda_j}=\prod_{a=1\atop{a\ne j}}\frac{\lambda_j-\lambda_a+ic}
 {\lambda_j-\lambda_a-ic},
 \ee
and all its solutions are real numbers. For the solution describing the ground state of the
model the spectral parameters $\{\lambda\}$  belong to an interval $[q,-q]$, where $q$ depends on the
coupling constant $c$ and the chemical potential $h$. The  system of twisted Bethe equations can be
written in the form
\be{NS-tBE}
 e^{iL(z_j-\frac{i\beta}L)}=\prod_{a=1\atop{a\ne
 j}}\frac{z_j-z_a+ic}{z_j-z_a-ic},\qquad \kappa=e^\beta.
 \ee
Therefore all the roots of the twisted Bethe equations belong to the line $\mathbb{R}+i\beta/L$.

The equation \eq{GME-1} takes the form
\be{NS-mult-int}
 \langle e^{\beta Q_1}\rangle=\frac{(-1)^N}{N!}\oint\limits_{\Gamma(\{\lambda\})}\prod_{j=1}^N
\left( \frac{dz_j}{2\pi i}\cdot e^{i(L/2-x)(z_j-\lambda_j)}\right)
 \cdot
 \frac{\Bigl[\det_N\Omega_\kappa(\{z\},\{\lambda\}|\{z\})\Bigr]^2}
 {\prod\limits_{j=1}^N{\cal Y}_\kappa(z_j|\{z\})
 \det_N\frac{\partial{\cal Y}(\lambda_j|\{\lambda\})}{\partial\lambda_k}}.
 \ee
Let $\beta<0$.
We will now transform this representation by several moves of the integration contour.
Note that the integrand vanishes when $|z_j|\to\infty$.
Therefore, one can evaluate the integral \eqref{NS-mult-int} by the residues at the poles
of the integrand outside the integration contour. These poles correspond to the solutions
of the twisted Bethe equations \eqref{NS-tBE}, which lie on the line $\mathbb{R}+i\beta/L$.
Thus, as soon as we have moved the contour to these roots, we can replace one of
$\det\Omega_\kappa$ by $\det\Omega$ due to \eq{OOk}, and we have
\begin{multline}\label{NS-int-TBE}
 \langle e^{\beta Q_1}\rangle=\frac{1}{N!}\prod_{j=1}^N
 \left(\int\limits_{\mathbb{R}+i\beta/L-i0}-\int\limits_{\mathbb{R}+i\beta/L+i0}\right)
 \frac{dz_j}{2\pi i}\cdot\prod_{j=1}^Ne^{-ix(z_j-\lambda_j)}
\nons
 \times
 \frac{\det_N\Omega_\kappa(\{z\},\{\lambda\}|\{z\})
 \det_N\Omega(\{\lambda\},\{z\}|\{\lambda\})}
 {\prod\limits_{j=1}^N{\cal Y}_\kappa(z_j|\{z\})
 \det_N\frac{\partial{\cal Y}(\lambda_j|\{\lambda\})}{\partial\lambda_k}}.
 \end{multline}
The integrals over $\mathbb{R}+i\beta/L-i0$ vanish due to the exponential decreasing of the factors
$\exp(-iz_jx)$ in the lower half plane. Thus, we obtain
\begin{multline}\label{NS-int-TBE-1}
 \langle e^{\beta Q_1}\rangle=\frac{(-1)^N}{N!}
 \int\limits_{\mathbb{R}+i\beta/L+i0}
 \prod_{j=1}^N\frac{dz_j}{2\pi i}\cdot\prod_{j=1}^Ne^{-ix(z_j-\lambda_j)}
\nons
 \times
 \frac{\det_N\Omega_\kappa(\{z\},\{\lambda\}|\{z\})
 \det_N\Omega(\{\lambda\},\{z\}|\{\lambda\})}
 {\prod\limits_{j=1}^N{\cal Y}_\kappa(z_j|\{z\})
 \det_N\frac{\partial{\cal Y}(\lambda_j|\{\lambda\})}{\partial\lambda_k}}.
 \end{multline}
It remains to move the contour up, through the real axis. Note that the matrix
$\Omega(\{\lambda\},\{z\}|\{\lambda\})$ has no poles at $z_j=\lambda_k$, therefore we cross only the
simple poles of $\det_N\Omega_\kappa(\{z\},\{\lambda\}|\{z\})$. Taking into account that
 \begin{multline}\label{res}
 \Res\left.\det_N\Omega_\kappa(\{z\},\{\lambda\}|\{z\})\right|_{z_a=\lambda_b}=(-1)^{a+b}
 {\cal Y}_\kappa(\lambda_b|\{z\ne z_a\},\lambda_b)\\
 \times\det_{N-1}\Omega_\kappa(\{z\ne z_a\},\{\lambda\ne\lambda_b\}|\{z\ne z_a\},\lambda_b),
 \end{multline}
we arrive at the sum over partitions
\begin{multline}\label{NS-int-part}
 \langle e^{\beta Q_1}\rangle=\sum_{n=0}^N\frac{(-1)^N}{n!}\sum_{\alpha\cup\bar\alpha\atop{\#\{\alpha\}=n}}
 \int\limits_{\mathbb{R}+i0}
 \prod_{j=1}^n\frac{dz_j}{2\pi i}\cdot\prod_{j=1}^ne^{-ixz_j}\prod_{a\in\alpha}e^{ix\lambda_a}
\nons
 \times
 \frac{\det_n\Omega_\kappa(\{z\},\{\lambda\}_\alpha|\{z\}\cup\{\lambda\}_{\bar\alpha})
 \det_N\Omega(\{\lambda\},\{z\}\cup\{\lambda\}_{\bar\alpha}|\{\lambda\})}
 {\prod\limits_{j=1}^n{\cal Y}_\kappa(z_j|\{z\}\cup\{\lambda\}_{\bar\alpha})
 \det_N\frac{\partial{\cal Y}(\lambda_j|\{\lambda\})}{\partial\lambda_k}}.
 \end{multline}
The representation \eq{NS-int-part} is the complete analog of the
one obtained for the $XXZ$ chain (see, for example,
\cite{KitMT00,KitMST02a,KitMST05a,KitMST05b}). Therefore the
remaining steps of the computation, in particular the
thermodynamic limit, are the same as the one's in the $XXZ$ model.
We refer the reader to these cited papers for the details and give
here the final result for the ground state expectation value of
the operator $e^{\beta Q_1}$ in the thermodynamic limit
\begin{multline}\label{answer}
\langle e^{\beta Q_1}\rangle= \sum_{n=0}^{\infty}\frac{1}{(n!)^2}\int\limits_{-q}^{q} d^n\lambda
\int\limits_{\mathbb{R}+i0}\prod_{j=1}^{n} \frac{dz_j}{2\pi i}\cdot \prod_{a,b=1}^n\frac{
(\lambda_a-z_b-ic)(z_b-\lambda_a-ic)} {(\lambda_a-\lambda_b-ic)(z_a-z_b-ic)}
          \\
\times \prod_{b=1}^ne^{-ix(z_b-\lambda_b)}\; \det_n\tilde M(\{\lambda\},\{z\})
\cdot\det_n[\tilde\rho(\lambda_j,z_k)].
\end{multline}
Here
 \be{tiM}
 \tilde M_{jk}(\{\lambda\},\{z\})=t(z_k,\lambda_j)+e^{\beta}
 t(\lambda_j,z_k)
 \prod_{a=1}^n\frac{h(\lambda_a,\lambda_j)h(\lambda_j,z_a)}{h(\lambda_j,\lambda_a)h(z_a,\lambda_j)},
 \ee
and the function $\tilde\rho(\lambda,z)$ solves the integral equation
\be{Int-eq}
 2\pi\tilde\rho(\lambda,z)-\int\limits_{-q}^q
 \frac{2c\tilde\rho(\mu,z)\,d\mu}{(\lambda-\mu)^2+c^2}=
 \frac{-\ c}{(\lambda-z)(\lambda-z-ic)}.
\ee

Observe that the series \eq{answer} is completely analogous to the one obtained for the $XXZ$ chain in
\cite{KitMST02a}. The function $\tilde\rho(\lambda,z)$ is just the rational limit of the corresponding
quantity for the $XXZ$ model. It is a universal function depending essentially on the $R$-matrix.

The series \eq{answer} is especially effective in the large coupling limit $c\to\infty$. Indeed, one can
see that only the first $m+2$ terms of the above series contribute to the $m-th$ order in $1/c$-expansion
of the density-density correlation function. In particular it reproduces the known results at the free
fermion point $c=\infty$ (see e.g. \cite{BogIK93L} and references therein).


\section*{Conclusion}

We have shown that the analog of the master equation for correlation functions  exists for a wide class of
integrable systems described by the $R$-matrix of the six-vertex model. Our method applies, in particular,
to the Quantum Nonlinear Schr\"odinger equation and the lattice Sine-Gordon model. Like in the case of the
$XXZ$ chain, this generalized master equation provides the direct analytical link between multiple
integral representations for the correlation functions and their expansions over form factors. Using the
generalized master equation one can also obtain various integral formulas for the correlation functions in
the thermodynamic limit. In the forthcoming publications we are planning to give such integral
representations, which are convenient for the asymptotic analysis of correlation functions.


\section*{Acknowledgments}
J. M. M., N. S. and V. T. are supported by CNRS.
N. K., K. K., J. M. M. and V. T. are supported by the European
network EUCLID-HPRNC-CT-2002-00325 and by the ANR programm GIMP ANR-05-BLAN-0029-01.
N.S. is supported by the French-Russian Exchange Program, the
Program of RAS Mathematical Methods of the Nonlinear Dynamics, RFBR-05-01-00498, Scientific Schools
672.2006.1. N. K and N. S. would like to thank the Theoretical Physics group of the Laboratory of
Physics at ENS Lyon for hospitality, which makes this collaboration possible.


\appendix

\section{Proof of the determinant representation for the scalar pro\-duct}
Consider an auxiliary function $G^{(n)}_N$ depending for given $n$ and $N$ ($0\le n\le N$) on three sets
of complex parameters $\{\xi_1,\dots,\xi_n\}$, $\{\nu_1,\dots,\nu_{N-n}\}$ and
$\{\lambda_1,\dots,\lambda_N\}$
\ba{defG}
 &&{\dis\hspace{-12mm}
 G^{(n)}_N(\{\xi\},\{\nu\},\{\lambda\})=\sum_{\alpha\cup\bar\alpha
\atop{\#\{\alpha\}=n}}(-1)^{ [{P}(\alpha)]}
 \det_{j\in\alpha\atop{k=1,\dots,n}}t(\lambda_j,\xi_k)
 \cdot\det_{j\in\bar\alpha\atop{k=1,\dots,N-n}} t(\nu_k,\lambda_j) }\non
 &&{\dis\hspace{-2mm}
 \times\left\{
 \prod_{a=1}^n\prod_{b=1}^{N-n}h(\nu_b,\xi_a)
 \prod_{a\in\alpha}\prod_{b\in\bar\alpha}
 h(\lambda_b,\lambda_a)- \prod_{a=1}^n\prod_{b\in\bar\alpha} h(\lambda_b,\xi_a)\prod_{a=1}^{N-n}\prod_{b\in\alpha}
  h(\nu_a,\lambda_b)\right\}}\non
 &&{\dis\hspace{62mm}
 \times\prod_{a=1}^n\prod_{b\in\alpha}h(\lambda_b,\xi_a)
 \prod_{a=1}^{N-n}\prod_{b\in\bar\alpha}h(\nu_a,\lambda_b).
 }
\ea
Here the sum is taken with respect to all partitions of the set $\{\lambda\}$ into two disjoint subsets
$\{\lambda\}=\{\lambda\}_{\alpha}\cup\{\lambda\}_{\bar\alpha}$, such that $\#\{\alpha\}=n$.

\begin{lemma}\label{Zero}
For arbitrary complex numbers $\{\xi_1,\dots,\xi_n\}$, $\{\nu_1,\dots,\nu_{N-n}\}$ and
$\{\lambda_1,\dots,\lambda_N\}$
\be{zero}
 G^{(n)}_N(\{\xi\},\{\nu\},\{\lambda\})\equiv0. \ee
\end{lemma}

{\sl Proof}.  For $N=1$ the statement of the lemma is evident, since either $n=0$ or $N-n=0$. Let
$G^{(n)}_{N-1}(\{\xi\},\{\nu\},\{\lambda\})=0$. Consider the properties of $G^{(n)}_N$ as a function of
$\lambda_1,\dots,\lambda_N$, with other parameters fixed. Obviously $G^{(n)}_N=0$ as soon as
$\lambda_j=\lambda_k$ ($j,k=1,\dots,N$, $j\ne k$). Hence, the combination
\be{combi}
 \tilde G^{(n)}_N(\{\xi\},\{\nu\},\{\lambda\})=
\frac{G^{(n)}_N(\{\xi\},\{\nu\},\{\lambda\})} {\prod\limits_{k<j}\sinh(\lambda_j-\lambda_k)}
 \ee
has no poles at $\lambda_j=\lambda_k$.  Therefore the only possible singularities of $\tilde G^{(n)}_N$
are the poles at $\lambda_j=\xi_k$ and $\lambda_j=\nu_k$.  It is easy to see that the residues in these
poles are proportional to $\tilde G^{(n)}_{N-1}$, for instance,
\ba{resid}
 &&{\dis\hspace{2mm}
\left.\Res\tilde G^{(n)}_N(\{\xi\},\{\nu\},\{\lambda\}) \right|_{\lambda_1=\xi_1}= \tilde
G^{(n)}_{N-1}(\{\xi\ne\xi_1\},\{\nu\},\{\lambda\ne\lambda_1\})}\non
 &&{\dis\hspace{2mm}
\times\sinh\eta\prod_{b=2}^N\frac{\sinh(\lambda_b-\xi_1+\eta)} {\sinh(\lambda_b-\xi_1)}\prod_{a=1}^n
 h(\xi_1,\xi_a)
 \prod_{a=1}^{N-n}h(\nu_a,\xi_1).}
 \ea
Due to the induction assumption these residues vanish. Thus, $\tilde G^{(n)}_N$ is a holomorphic function
of each $\lambda_j$ in whole complex plane. It remains to observe that $\tilde
G^{(n)}_N(\{\xi\},\{\nu\},\{\lambda\})$ is an $i\pi$-anti-periodic function of the
parameters $\{\lambda\}$ and that it exponentially decreases at $|\Re\lambda_j|\to\infty$.
Hence, $G^{(n)}_N(\{\xi\},\{\nu\},\{\lambda\})=0$.
\qed

Consider now the equation \eq{Gen-case}. Let one of the states, for instance
$\prod_{j=1}^NB(\lambda_j)|0\rangle$, be an eigenstate of the twisted transfer-matrix. Then, due to the
twisted Bethe equations \eq{TTMBE_Y}, one can express $r(\lambda_j)$ as
 \be{BE}
 r(\lambda_j)=\kappa (-1)^{N-1}\prod_{a=1}^N\frac{h(\lambda_j,\lambda_a)}{h(\lambda_a,\lambda_j)}.
 \ee
Substituting this into \eq{Gen-case} we obtain
 \begin{multline}\label{Gen-case-A}
 \langle0|\prod_{j=1}^N\frac{C(\mu_j)}{d(\mu_j)}
 \prod_{j=1}^N\frac{B(\lambda_j)}{d(\lambda_j)}|0\rangle=\prod_{j>k}^N
 \frac1{\sinh(\lambda_j-\lambda_k)\sinh(\mu_k-\mu_j)}\nons
 \times\sum_{\alpha\cup\bar\alpha\atop{\gamma\cup\bar\gamma}}(-1)^{[P(\alpha)]+[P(\gamma)]}
 (-1)^{nN-n}\kappa^{N-n} \prod_{a\in\gamma}r(\mu_a)
 \det_{j\in\alpha\atop{k\in\gamma}}t(\lambda_j,\mu_k)
 \det_{j\in\bar\alpha\atop{k\in\bar\gamma}}t(\mu_k,\lambda_j)\nons
 \times
 \prod_{a\in\gamma}\prod_{b\in\alpha}h(\lambda_b,\mu_a)
 \prod_{a\in\bar\alpha}\prod_{b\in\bar\gamma}h(\mu_b,\lambda_a)
 \prod_{a\in\bar\alpha}\prod_{b\in\alpha}h(\lambda_a,\lambda_b)
 \prod_{a\in\gamma}\prod_{b\in\bar\gamma}h(\mu_b,\mu_a),
 \end{multline}
where $n=\#\{\alpha\}$. For given partitions $\{\gamma\}$ and $\{\bar\gamma\}$ one can set
$\{\mu\}_{\gamma}=\{\xi_1,\dots,\xi_n\}$,  $\{\mu\}_{\bar\gamma}=\{\nu_1,\dots,\nu_{N-n}\}$ and then apply
Lemma \ref{Zero}
 \begin{multline}\label{Gen-case-A1}
 \langle0|\prod_{j=1}^N\frac{C(\mu_j)}{d(\mu_j)}
 \prod_{j=1}^N\frac{B(\lambda_j)}{d(\lambda_j)}|0\rangle=\prod_{j>k}^N
 \frac1{\sinh(\lambda_j-\lambda_k)\sinh(\mu_k-\mu_j)}\nons
 \times\sum_{\alpha\cup\bar\alpha\atop{\gamma\cup\bar\gamma}}(-1)^{[P(\alpha)]+[P(\gamma)]}
 (-1)^{nN-n}\kappa^{N-n} \prod_{a\in\gamma}r(\mu_a)
 \det_{j\in\alpha\atop{k\in\gamma}}t(\lambda_j,\mu_k)
 \det_{j\in\bar\alpha\atop{k\in\bar\gamma}}t(\mu_k,\lambda_j)\nons
 \times
 \prod_{a\in\gamma}\prod_{b=1}^N h(\lambda_b,\mu_a)
 \prod_{a=1}^N\prod_{b\in\bar\gamma}h(\mu_b,\lambda_a).
  \end{multline}
It remains to use the Laplace formula for the determinant of the sum of two matrices and we arrive at
\eq{Sc-prod}.


\section{Identity for determinants }\label{id-det}
Let
 \be{hM}
 \hat G_{jk}=t(\mu_k,\lambda_j)\prod_{a=1}^Nh(\mu_k,\lambda_a)+\kappa
 t(\lambda_j,\mu_k)\prod_{a=1}^Nh(\lambda_a,\mu_k)
 \prod_{a=1}^N\frac{h(\mu_k,\mu_a)}{h(\mu_a,\mu_k)},
\ee
 \be{tM}
 \tilde G_{jk}=t(\mu_k,\lambda_j)\prod_{a=1}^Nh(\mu_a,\lambda_j)+\kappa
 t(\lambda_j,\mu_k)\prod_{a=1}^Nh(\lambda_j,\mu_a)
 \prod_{a=1}^N\frac{h(\lambda_a,\lambda_j)}{h(\lambda_j,\lambda_a)}.
 \ee

\begin{lemma}\label{Identity}
 For arbitrary complex numbers $\lambda_1,\dots,\lambda_N$ and $\mu_1,\dots,\mu_N$ the following identity holds
 \be{2-forms}
 \det_N \tilde G =\det_N \hat G.
 \ee
\end{lemma}

{\sl Proof.} The proof of this lemma is similar to the one of Lemma \ref{Zero}. For $N=1$ the statement of
the lemma is correct. Assume that it is still correct for $N-1$. Then, considering the equation
\eq{2-forms} in the points $\lambda_j=\mu_k$, we see that the corresponding residues are proportional to
the determinants of the size $N-1$, which are equal to each other due to the inductive assumption.

The only nontrivial step is to prove that the determinants of $\tilde G$ and $\hat G$  have no other
singularities except at the points $\lambda_j=\mu_k$. At the first sight one could expect that $\det\tilde
G$  might have simple poles at $h(\lambda_j,\lambda_k)=0$, and  $\det\hat G$  at $h(\mu_j,\mu_k)=0$. It is
easy to see, however, that the corresponding residues vanish. Indeed, let, for example,
$h(\lambda_1,\lambda_2)=0$. Then the first line of the matrix $\tilde G$ becomes singular
 \be{tM1}
 \tilde G_{1k}\sim \kappa
 t(\lambda_1,\mu_k)\prod_{a=1}^Nh(\lambda_1,\mu_a)
 \prod_{a=1}^N\frac{h(\lambda_a,\lambda_1)}{h(\lambda_1,\lambda_a)}.
 \ee
On the other hand one has for the second line of $\tilde G$
 \be{tM2}
 \tilde G_{2k}\sim t(\mu_k,\lambda_2)\prod_{a=1}^Nh(\mu_a,\lambda_2).
 \ee
Taking into account that $t(\mu_k,\lambda_2)=t(\lambda_1,\mu_k)$ at $h(\lambda_1,\lambda_2)=0$, we see
that the first and the second lines of $\tilde G$ are proportional to each other. Hence, the corresponding
residue vanishes. Similarly one can prove that $\det\hat G$ has no poles at $h(\mu_j,\mu_k)=0$.
\qed

The identity \eq{Identity} allows one to prove directly the equation \eq{OOk} if the parameters
$\{\lambda\}$ (respectively $\{\mu\}$) are solutions of twisted Bethe equations \eq{TTMBE_Y} (respectively
ordinary Bethe equations with $\kappa=1$).


\section{Proof of the theorem \ref{M-theor}}\label{PT}

We start with the equation \eq{IK}. Substituting \eq{Gen-case} for each scalar product we obtain a sum
over $4$ partitions of the set $\{\lambda\}$ and $4$ partitions of the set $\{\mu\}$:
 \begin{multline}
 \Qk=\sum_{\alpha_{+}\cup\alpha_{-}\cup\bar\alpha_{+}\cup\bar\alpha_{-}
 \atop{\gamma_{+}\cup\gamma_{-}\cup\bar\gamma_{+}\cup\bar\gamma_{-}}}
  \prod_{a\in\gamma_{+}\cup\bar\gamma_{-}}l(\mu_a)
 \prod_{a\in\alpha_{+}\cup\bar\alpha_{-}}m(\lambda_a)
 \prod_{a\in\bar\gamma_{+}}r(\mu_a)\prod_{a\in\alpha_{-}}r(\lambda_a)\nons
 \times\kappa^{n_{+}+n_{-}}
 \det_{j\in\alpha_{+}\atop{k\in\gamma_{+}}}t(\lambda_j,\mu_k)
 \det_{j\in\alpha_{-}\atop{k\in\gamma_{-}}}t(\mu_k,\lambda_j)
 \det_{j\in\bar\alpha_{+}\atop{k\in\bar\gamma_{+}}}t(\lambda_j,\mu_k)
 \det_{j\in\bar\alpha_{-}\atop{k\in\bar\gamma_{-}}}t(\mu_k,\lambda_j)\nons
 \times
 \prod_{a\in\gamma_{+}}\prod_{b\in\alpha_{+}}h(\lambda_b,\mu_a)
 \prod_{a\in\alpha_{-}}\prod_{b\in\gamma_{-}}h(\mu_b,\lambda_a)
 \prod_{a\in\alpha_{-}}\prod_{b\in\alpha_{+}}h(\lambda_b,\lambda_a)
 \prod_{a\in\gamma_{+}}\prod_{b\in\gamma_{-}}h(\mu_b,\mu_a)\nons
 \times
 \prod_{a\in\bar\gamma_{+}}\prod_{b\in\bar\alpha_{+}}h(\lambda_b,\mu_a)
 \prod_{a\in\bar\alpha_{-}}\prod_{b\in\bar\gamma_{-}}h(\mu_b,\lambda_a)
 \prod_{a\in\bar\alpha_{-}}\prod_{b\in\bar\alpha_{+}}h(\lambda_b,\lambda_a)
 \prod_{a\in\bar\gamma_{+}}\prod_{b\in\bar\gamma_{-}}h(\mu_b,\mu_a)\nons
 \times (-1)^{[P(\alpha)]+[P(\gamma)]}
 \prod_{a\in\alpha_{+}\cup\alpha_{-}}\prod_{b\in\bar\alpha_{+}\cup\bar\alpha_{-}}
 h(\lambda_b,\lambda_a)
 \prod_{a\in\gamma_{+}\cup\gamma_{-}}\prod_{b\in\bar\gamma_{+}\cup\bar\gamma_{-}}
 h(\mu_a,\mu_b).\label{matr-el}
 \end{multline}
Here, as usual, $P(\alpha)$ is the permutation acting as
$P(\{\alpha_{+}\},\{\alpha_{-}\},\{\bar\alpha_{+}\},\{\bar\alpha_{-}\})=1,\dots,N$ and the permutation
$P(\gamma)$ acts in a similar way. The numbers of elements in the subsets $\{\alpha_{+}\}$ and
$\{\alpha_{-}\}$ are denoted as $n_{+}$ and $n_{-}$ respectively.

The sets $\{\mu\}$ and $\{\lambda\}$ in \eq{matr-el} are still generic complex numbers. If the set
$\{\lambda\}$ satisfies Bethe equations \eq{TTMBE_Y}  with $\kappa=1$, then we can express $r(\lambda_a)$
similarly to \eq{BE}. Denoting
 \be{new-part}
 \begin{array}{l}
 \alpha_0=\alpha_{+}\cup\bar\alpha_{-},\qquad
 \bar\alpha_0=\bar\alpha_{+}\cup\alpha_{-},\\
 \gamma_0=\gamma_{+}\cup\bar\gamma_{-},\qquad
 \bar\gamma_0=\bar\gamma_{+}\cup\gamma_{-},
 \end{array}
 \ee
we obtain
 \begin{multline}\label{Mat-el-bet}
  \Qk=\sum_{\alpha_{+}\cup\alpha_{-}\cup\bar\alpha_{+}\cup\bar\alpha_{-}
 \atop{\gamma_{+}\cup\gamma_{-}\cup\bar\gamma_{+}\cup\bar\gamma_{-}}}
 (-1)^{[P(\alpha)]+[P(\gamma)]}
 \prod_{a\in\gamma_0}l(\mu_a)
 \prod_{a\in\alpha_0}m(\lambda_a)\cdot \kappa^{n_{+}}
  \nons
 \times
 \det_{j\in\alpha_{+}\atop{k\in\gamma_{+}}}t(\lambda_j,\mu_k)
 \det_{j\in\bar\alpha_{-}\atop{k\in\bar\gamma_{-}}}t(\mu_k,\lambda_j)
 \prod_{a\in\bar\alpha_0}\prod_{b\in\alpha_0}
 h(\lambda_a,\lambda_b)
 \prod_{b\in\gamma_{-}}\prod_{a\in\gamma_0}
 h(\mu_b,\mu_a)
 \prod_{b\in\bar\gamma_{+}}\prod_{a\in\gamma_0}
 h(\mu_a,\mu_b)
 \nons
 \times
 \prod_{a\in\gamma_{+}}\prod_{b\in\alpha_{+}}h(\lambda_b,\mu_a)
 \prod_{a\in\bar\alpha_{-}}\prod_{b\in\bar\gamma_{-}}h(\mu_b,\lambda_a)
 \prod_{a\in\bar\alpha_{-}}\prod_{b\in\alpha_{+}}h(\lambda_a,\lambda_b)
 \prod_{a\in\gamma_{+}}\prod_{b\in\bar\gamma_{-}}h(\mu_a,\mu_b)
 \nons
 \times
 \kappa^{n_{-}} (-1)^{(N-1)n_{-}}\prod_{a\in\bar\gamma_{+}}r(\mu_a)
 \det_{j\in\alpha_{-}\atop{k\in\gamma_{-}}}t(\mu_k,\lambda_j)
 \det_{j\in\bar\alpha_{+}\atop{k\in\bar\gamma_{+}}}t(\lambda_j,\mu_k)
 \nons
 \times
 \prod_{a\in\alpha_{-}}\prod_{b\in\gamma_{-}}h(\mu_b,\lambda_a)
 \prod_{a\in\bar\gamma_{+}}\prod_{b\in\bar\alpha_{+}}h(\lambda_b,\mu_a)
 \prod_{a\in\alpha_{-}}\prod_{b\in\bar\alpha_{+}}h(\lambda_a,\lambda_b)
 \prod_{a\in\gamma_{-}}\prod_{b\in\bar\gamma_{+}}h(\mu_a,\mu_b).
\end{multline}
The sets $\{\lambda\}_{\alpha_{-}}$ and $\{\lambda\}_{\bar\alpha_{+}}$ enter separately only the two last
lines of \eq{Mat-el-bet}. Hence, we can sum up over these partitions (at others partitions fixed) and
apply Lemma \ref{Zero}:
 \begin{multline}\label{interm-1}
 \sum_{\alpha_{-}\cup\bar\alpha_{+}}(-1)^{[P(\alpha)]}
 \det_{j\in\alpha_{-}\atop{k\in\gamma_{-}}}t(\mu_k,\lambda_j)
 \det_{j\in\bar\alpha_{+}\atop{k\in\bar\gamma_{+}}}t(\lambda_j,\mu_k)\nons
 \times
 \prod_{a\in\alpha_{-}}\prod_{b\in\gamma_{-}}h(\mu_b,\lambda_a)
 \prod_{a\in\bar\gamma_{+}}\prod_{b\in\bar\alpha_{+}}h(\lambda_b,\mu_a)
 \prod_{a\in\alpha_{-}}\prod_{b\in\bar\alpha_{+}}h(\lambda_a,\lambda_b)
 \prod_{a\in\gamma_{-}}\prod_{b\in\bar\gamma_{+}}h(\mu_a,\mu_b)\nons
 =\sum_{\alpha_{-}\cup\bar\alpha_{+}}(-1)^{[P(\alpha)]}
 \det_{j\in\alpha_{-}\atop{k\in\gamma_{-}}}t(\mu_k,\lambda_j)
 \det_{j\in\bar\alpha_{+}\atop{k\in\bar\gamma_{+}}}t(\lambda_j,\mu_k)\nons
 \times
 \prod_{b\in\bar\gamma_{+}}\prod_{a\in\bar\alpha_0}h(\lambda_a,\mu_b)
 \prod_{b\in\gamma_{-}}\prod_{a\in\bar\alpha_0}h(\mu_b,\lambda_a).
 \end{multline}
Substituting this into \eq{Mat-el-bet} we see that the sum over the partitions
$\{\alpha_{-}\},\{\bar\alpha_{+}\},\{\gamma_{-}\},\{\bar\gamma_{+}\}$ gives exactly the Laplace
decomposition of the determinant. Thus, we obtain
 \begin{multline}\label{Mat-el-bet-1}
  \Qk=\sum_{\alpha_{+}\cup\bar\alpha_{-}\cup\bar\alpha_{0}
 \atop{\gamma_{+}\cup\bar\gamma_{-}\cup\bar\gamma_0}}
  \prod_{a\in\gamma_0}l(\mu_a) \prod_{a\in\alpha_0}m(\lambda_a)\cdot \kappa^{n_{+}}
 \cdot
 \det_{j\in\alpha_{+}\atop{k\in\gamma_{+}}}t(\lambda_j,\mu_k)
 \det_{j\in\bar\alpha_{-}\atop{k\in\bar\gamma_{-}}}t(\mu_k,\lambda_j)\nons
 \times(-1)^{[P(\alpha)]+[P(\gamma)]}
 \det_{j\in\bar\alpha_0\atop{k\in\bar\gamma_0}}\left[r(\mu_k)t(\lambda_j,\mu_k)
 \prod_{a\in\bar\alpha_0}h(\lambda_a,\mu_k)\prod_{a\in\gamma_0}h(\mu_a,\mu_k)
 \right.\nons
 -\left.  \kappa(-1)^Nt(\mu_k,\lambda_j)
  \prod_{a\in\bar\alpha_0}h(\mu_k,\lambda_a)\prod_{a\in\gamma_0}h(\mu_k,\mu_a)\right]
  \cdot
  \prod_{a\in\bar\alpha_0}\prod_{b\in\alpha_0}
  h(\lambda_a,\lambda_b)\nons
  \times
  \prod_{a\in\gamma_{+}}\prod_{b\in\alpha_{+}}h(\lambda_b,\mu_a)
 \prod_{a\in\bar\alpha_{-}}\prod_{b\in\bar\gamma_{-}}h(\mu_b,\lambda_a)
 \prod_{a\in\bar\alpha_{-}}\prod_{b\in\alpha_{+}}h(\lambda_a,\lambda_b)
 \prod_{a\in\gamma_{+}}\prod_{b\in\bar\gamma_{-}}h(\mu_a,\mu_b).
   \end{multline}
Now, for arbitrary fixed  partitions $\bar\alpha_0$ and $\bar\gamma_0$, we can sum up with respect to
the remaining partitions. Using again Lemma \ref{Zero} we have
 \begin{multline}\label{interm-2}
 \sum_{\alpha_{+}\cup\bar\alpha_{-}
 \atop{\gamma_{+}\cup\bar\gamma_{-}}}
 (-1)^{[P(\alpha)]+[P(\gamma)]} \kappa^{n_{+}} \cdot
 \det_{j\in\alpha_{+}\atop{k\in\gamma_{+}}}t(\lambda_j,\mu_k)
 \det_{j\in\bar\alpha_{-}\atop{k\in\bar\gamma_{-}}}t(\mu_k,\lambda_j)
 \cdot
 \prod_{a\in\gamma_+}\prod_{b\in\gamma_0}\frac{h(\mu_a,\mu_b)}{h(\mu_b,\mu_a)}\nons
   \times
  \prod_{a\in\gamma_{+}}\prod_{b\in\alpha_{+}}h(\lambda_b,\mu_a)
 \prod_{a\in\bar\alpha_{-}}\prod_{b\in\bar\gamma_{-}}h(\mu_b,\lambda_a)
 \prod_{a\in\bar\alpha_{-}}\prod_{b\in\alpha_{+}}h(\lambda_a,\lambda_b)
 \prod_{a\in\gamma_{+}}\prod_{b\in\bar\gamma_{-}}h(\mu_b,\mu_a)\nons
 =\sum_{\alpha_{+}\cup\bar\alpha_{-}
 \atop{\gamma_{+}\cup\bar\gamma_{-}}}
 (-1)^{[P(\alpha)]+[P(\gamma)]} \kappa^{n_{+}} \cdot
 \det_{j\in\alpha_{+}\atop{k\in\gamma_{+}}}t(\lambda_j,\mu_k)
 \det_{j\in\bar\alpha_{-}\atop{k\in\bar\gamma_{-}}}t(\mu_k,\lambda_j)
 \cdot
 \prod_{a\in\gamma_+}\prod_{b\in\gamma_0}\frac{h(\mu_a,\mu_b)}{h(\mu_b,\mu_a)}\nons
   \times
  \prod_{b\in\gamma_{+}}\prod_{a\in\alpha_{0}}h(\lambda_a,\mu_b)
 \prod_{b\in\bar\gamma_{-}}\prod_{a\in\alpha_{0}}h(\mu_b,\lambda_a)\nons
 =\det_{j\in\alpha_0\atop{k\in\gamma_0}}\left[
 t(\mu_k,\lambda_j)\prod_{a\in\alpha_{0}}h(\mu_k,\lambda_a)+\kappa
 t(\lambda_j,\mu_k)\prod_{a\in\alpha_{0}}h(\lambda_a,\mu_k)
\prod_{a\in\gamma_0}\frac{h(\mu_k,\mu_a)}{h(\mu_a,\mu_k)}\right].
   \end{multline}
Applying now Lemma \ref{Identity} we finally arrive at
 \begin{multline}\label{final-1}
  \Qk=\sum_{\alpha_0\cup\bar\alpha_{0}
 \atop{\gamma_0\cup\bar\gamma_0}}
 (-1)^{[P(\alpha)]+[P(\gamma)]}
 \prod_{a\in\gamma_0}l(\mu_a) \prod_{a\in\alpha_0}m(\lambda_a)
 \cdot\prod_{a\in\bar\alpha_0}\prod_{b\in\alpha_0}
  h(\lambda_a,\lambda_b)\nons
  \times\det_{j\in\alpha_0\atop{k\in\gamma_0}}\left[
 t(\mu_k,\lambda_j)\prod_{a\in\gamma_{0}}h(\mu_a,\lambda_j)+\kappa
 t(\lambda_j,\mu_k)\prod_{a\in\gamma_{0}}h(\lambda_j,\mu_a)
 \prod_{a\in\alpha_0}\frac{h(\lambda_a,\lambda_j)}{h(\lambda_j,\lambda_a)}\right]
 \nons
 \times\det_{j\in\bar\alpha_0\atop{k\in\bar\gamma_0}}\left[r(\mu_k)t(\lambda_j,\mu_k)
 \prod_{a\in\bar\alpha_0}h(\lambda_a,\mu_k)\prod_{a\in\gamma_0}h(\mu_a,\mu_k)
 \right.\nons
 -\left.  \kappa(-1)^Nt(\mu_k,\lambda_j)
  \prod_{a\in\bar\alpha_0}h(\mu_k,\lambda_a)\prod_{a\in\gamma_0}h(\mu_k,\mu_a)\right].
  \end{multline}

It is easy to see that this sum over partitions can be written as a single multiple integral
\begin{multline}\label{mult-int}
 \Qk=\frac1{N!}\oint\limits_{\Gamma(\{\mu\}\cup\{\lambda\})}\prod_{j=1}^N
 \frac{dz_j}{2\pi i}\prod_{j=1}^N\frac{l(z_j)l^{-1}(\lambda_j)d(z_j)}
 {d(\lambda_j)d(\mu_j)}\nons
 \times
 \frac{\det_N\Omega_\kappa(\{z\},\{\mu\}|\{z\})\det_N\Omega_\kappa(\{z\},\{\lambda\}|\{z\})}
 {\prod\limits_{j=1}^N{\cal Y}_\kappa(z_j|\{z\})}.
 \end{multline}
Here the integration contour $\Gamma(\{\mu\}\cup\{\lambda\})$ surrounds the poles of $\Omega_\kappa$ in
the points $\{\lambda\}$ and $\{\mu\}$ and no other pole of the integrand.



\end{document}